\documentclass{article}
\usepackage{slashed,amsmath,amssymb,enumitem}
\usepackage[papersize={8.5in,11in}]{geometry}
\begin{document}
\title{Can Electroweak Theory Without A Higgs Particle Be Renormalizable?}
\author{J. W. Moffat\\~\\
Perimeter Institute for Theoretical Physics, Waterloo, Ontario N2L 2Y5, Canada\\
and\\
Department of Physics and Astronomy, University of Waterloo, Waterloo,\\
Ontario N2L 3G1, Canada}
\maketitle
\begin{abstract}
Whether there exists a massive electroweak (EW) theory, without a Higgs spontaneous symmetry breaking mechanism, that is gauge invariant and renormalizable is investigated. A Stueckelberg formalism for massive $W$ and $Z$ bosons is used to derive a gauge invariant EW theory. Negative energy scalar fields that emerge from the gauge invariance of the Lagrangian are removed by invoking an indefinite metric in Hilbert space. A unitary S-matrix and a positive energy spectrum can be obtained by using the PT symmetric formulation of the pseudo-Hermitian Hamiltonian. The theory predicts that if for a system of particles the scalar boson energy $E_s < \mu=\lambda^{1/2}M_W$, where $\lambda$ is a gauge parameter and $M_W$ is the $W$ boson mass, then as $\lambda\rightarrow\infty$ the scalar boson mass $\mu=\lambda^{1/2}M_W$ tends to infinity. The theory is perturbatively renormalizable and does not violate longitudinally polarized $W_L W_L\rightarrow W_L W_L$ scattering in the energy range $E < \mu=\lambda^{1/2}M_W$ for which the scalar bosons have decoupled and they have an undetected mass. This means that with this scenario the EW theory can only be treated as an effective renomalizable theory and not as a UV complete theory.
\end{abstract}


\section{Introduction}

There is no experimental evidence that the Higgs particle exists. If the Higgs particle is not detected, then we must consider revising at a fundamental level the electroweak (EW) model of Weinberg and Salam~\cite{Weinberg,Salam,Aitchison}. This may require a revision of our ideas about QFT. A previously published EW theory without a Higgs particle and a quantum gravity theory~\cite{Moffat,Moffat2,Moffat3,Moffat4} were based on nonlocal interactions and the EW theory led to finite amplitudes and cross sections that can be tested at the LHC.

In this paper we consider whether another approach is also viable: can we construct a local, physically consistent and renormalizable EW model containing only the observed particles, namely, 12 quarks and leptons, the charged $W$ boson, the neutral $Z$ boson and the massless photon and gluon without the Higgs particle? The renormalizable theory should not violate unitarity for the tree graph calculation of $W_LW_L\rightarrow W_LW_L$ longitudinally polarized scattering above an energy of 1-2 TeV.

In the following, we will explore an EW model based on a gauge invariant action with local interactions. The gauge invariance of the Lagrangian for massive $W$ and $Z$ vector boson fields is obtained using a Stuekelberg formalism~\cite{Stueckelberg,Goto,Burnel}. The gauge invariance of the Lagrangian leads to the existence of Ward-Takahashi-Slavnov-Taylor~\cite{Ward} identities and a conserved current. This guarantees a renormalizable EW theory, provided the scalar fields with negative norm in the Lagrangian decouple at high energies rendering a scalar spin-0 boson undetectable at present accelerator energies. In general, for a scalar field with negative norm, it is necessary to introduce an indefinite metric in Hilbert space~\cite{Pauli}. This renders the S matrix non-unitary and the Hamiltonian is non-Hermitian. This problem can be resolved by the Hamiltonian being PT symmetric and invoking the methods developed by Bender and collaborators~\cite{Bender,Bender2}.

\section{The Gauge Invariant Electroweak Lagrangian}

The theory introduced here is based on a Lagrangian that includes leptons and quarks with the color degree
of freedom of the strong interaction group $SU_c(3)$. We shall use the metric convention, $\eta_{\mu\nu}={\rm diag}(+1,-1,-1,-1)$, and set $\hbar=c=1$. Our gauge invariant EW Lagrangian is of the form:
\begin{align}
\label{GaugeEWLagrangian}
{\cal L}_{\rm EW}=\sum_f {\bar f}i\slashed{D}^W f+\sum_f{\bar f}i\slashed{D}^Z f
+\sum_f{\bar f}i\slashed{D}^A f-\frac{1}{2}W^{+\mu\nu}W^{-}_{\mu\nu}-\frac{1}{4}Z_{\mu\nu}Z^{\mu\nu}-\frac{1}{4}F^{\mu\nu}F_{\mu\nu}\nonumber\\
+\frac{1}{2}(M_ZZ_\mu-\partial_\mu\beta)(M_Z Z^\mu-\partial^\mu\beta)
+(M_WW^+_\mu-P\partial_\mu\sigma)(M_WW^{-\mu}-P\partial^{\mu}\sigma)+{\cal L}_{m_f},
\end{align}
where $\beta$ and $\sigma$ are scalar gauge fields and $P$ is a function of $\sigma$. The Lagrangian (\ref{GaugeEWLagrangian}) is invariant under the infinitesimal gauge transformations
\begin{equation}
Z_\mu\rightarrow Z_\mu+\partial_\mu\nu,\quad \beta\rightarrow \beta+M_Z\nu,
\end{equation}
and
\begin{equation}
W_\mu\rightarrow W_\mu+{\cal D}^W_\mu\chi,\quad \sigma\rightarrow \sigma+Q\chi.
\end{equation}
Here, $Q$ is an unknown function of $\chi$, ${\cal D}^W_\mu$ is a covariant differential operator and $\slashed{D}^{W,Z,A}=\gamma^\mu D^{W,Z,A}_\mu$. We have $W^+_\mu=(W^-_\mu)^\dagger$ where
\begin{equation}
W^+_\mu=\frac{1}{\sqrt{2}}(W^1_\mu-iW^2_\mu),\quad W^-_\mu=\frac{1}{\sqrt{2}}(W^1_\mu+iW^2_\mu),
\end{equation}
and
\begin{equation}
\label{WZequation}
W^+_{\mu\nu}=D^W_\mu W^+_\nu-D^W_\nu W^+_\mu,\quad Z_{\mu\nu}=\partial_\mu Z_\nu -\partial_\nu Z_\mu,
\end{equation}
where $W^+_{\mu\nu}=(W^-_{\mu\nu})^\dagger$. The non-Abelian field $W^a_{\mu\nu}$ is given by
\begin{equation}
W^a_{\mu\nu}=\partial_\mu W^a_\nu-\partial_\nu W^a_\mu+gf^{abc}W^b_\mu W^c_\nu.
\end{equation}
We have $W_\mu=T^aW^a_\mu$, $T^a$ are the generators of the group $SU(2)$ and $f^{abc}=\epsilon^{abc}$ are the structure constants.
Moreover, the photon field $F_{\mu\nu}$ is
\begin{equation}
\label{Fequation}
F_{\mu\nu}=\partial_\mu A_\nu-\partial_\nu A_\mu,
\end{equation}
and
\begin{equation}
D^W_\mu=\partial_\mu-igW_\mu,\quad D^Z_\mu=\partial_\mu-ig'Z_\mu,\quad D^A_\mu=\partial_\mu-ieA_\mu.
\end{equation}
The fermion sums in (\ref{GaugeEWLagrangian}) run over all quark and lepton fields. All the fields are {\it local fields} that satisfy microcausality.

The fermion mass Lagrangian is
\begin{equation}
\label{fermionmass}
{\cal L}_{m_f}=-\sum_{\psi_L^i,\psi_R^j}m_{ij}^f(\bar\psi_L^i\psi_R^j + \bar\psi_R^i\psi_L^j),
\end{equation}
where $\psi_{L,R}=P_{L,R}\psi$, $P_{L,R}=\frac{1}{2}(1\mp\gamma_5)$ and $m_{ij}^f$ denotes the fermion masses. Eq. (\ref{fermionmass}) can incorporate massive neutrinos and their flavor oscillations. We have not included in the Lagrangian $(\ref{GaugeEWLagrangian})$ the standard scalar field Higgs contribution:
\begin{equation}
\label{scalarLagrangian}
{\cal L}_\phi=\large\vert(i\partial_\mu - gT^aW^a_\mu - g'\frac{Y}{2}B_\mu)\phi\large\vert^2-V(\phi),
\end{equation}
where $\vert...\vert^2=(...)^\dagger(...)$. Moreover,
\begin{equation}
\label{potential}
V(\phi)=\mu_H^2\phi^\dagger\phi +\lambda_H(\phi^\dagger\phi)^2,
\end{equation}
where $B_\mu$ is the neutral vector boson that couples to weak hypercharge, $\mu^2_H < 0$ and $\lambda_H > 0$. The $W$ and $Z$ masses are the experimental masses. We do not begin with a massless Lagrangian and then break the $SU(2)$ symmetry through a spontaneous symmetry breaking of the vacuum.

The electromagnetic current is
\begin{equation}
J_\mathrm{em}^\mu=J^{3\mu}+J_Y^\mu=\sum_fq_f{\bar f}\gamma^\mu f,
\end{equation}
and
\begin{equation}
J^{3\mu}=\sum_fT^3_f{\bar f}\gamma^\mu f,
\end{equation}
where $q_f$, $Y$ and $T^3$ denote the fermions' electric charge, hypercharge and weak isospin, respectively. The neutral current is given by
\begin{equation}
J^{{\rm NC}\mu}=J^{3\mu}-\sin^2\theta_wJ_\mathrm{em}^\mu,
\end{equation}
while the charge current is
\begin{equation}
J^{\mathrm{C}\mu}={\bar u}_i\gamma^\mu\frac{1}{2}(1-\gamma^5)M_{ij}^{\rm CKM}d_j+{\bar\nu}_i\gamma^\mu\frac{1}{2}(1-\gamma^5)e_i,
\end{equation}
where $M_{ij}^{\rm CKM}$ denotes the Cabbibo-Kobayashi-Maskawa (CKM) matrix.

The fermion-boson interaction terms are given by
\begin{equation}
{\cal L}_{fV}=-\frac{g}{\sqrt{2}}(J_\mu^+W^{+\mu}+J_\mu^-W^{-\mu})-g\sin\theta_wJ_\mathrm{em}^\mu A_\mu-\frac{g}{\cos\theta_w}
J^{\mathrm{NC}\mu} Z_\mu.
\end{equation}
The vector three-point $WWZA$ self interaction Lagrangian is
\begin{equation}
{\cal L}_{WWV}=-ig[W^+_{\mu\nu}W^{-\mu}-W^{+\mu}W_{\mu\nu}^-)(A^\nu\sin\theta_w-Z^\nu\cos\theta_w)+W^-_\nu W^+_\mu(F^{\mu\nu}\sin\theta_w-Z^{\mu\nu}\cos\theta_w)].
\end{equation}
The vector four-point self interaction Lagrangian is
\begin{align}
{\cal L}_{WWVV}&=-\frac{g^2}{4}\{[2W^+_\mu W^{-\mu}+(A_\mu\sin\theta_w-Z_\mu\cos\theta_w)^2]^2\nonumber\\
&-[W^+_\mu W^-_\nu+W^+_\nu W^-_\mu+(A_\mu\sin\theta_w-Z_\mu\cos\theta_w)(A_\nu\sin\theta_w-Z_\nu\cos\theta_w)]^2\}.
\end{align}

We have adopted the unification condition
\begin{equation}
e=g\sin\theta_w=g'\cos\theta_w.
\end{equation}

\section{The Charged Vector Boson Propagator and Renormalizability}

It was recognized from the beginnings of investigations of EW models that introducing massive charged gauge
bosons $W^\pm$ in the form of mass terms $M_W^2W^{+\mu}W^-_\mu$ into the Lagrangian, produces non-renormalizable divergences. When we calculate loop
diagrams with massive bosons in standard local QFT, we get for the amplitude:
\begin{equation}
\label{loopintegral}
{\rm Amplitude}=\int d^4p({\rm propagators})\cdots.
\end{equation}
For massive boson propagators of the form:
\begin{equation}
\label{masspropagator}
D^{V\mu\nu}(p^2)=\frac{i\biggl(-\eta^{\mu\nu} + \frac{p^\mu p^\nu}{M^2_V}\biggr)}{p^2-M^2_V+i\epsilon},
\end{equation}
we have for large $p^2$:
\begin{equation}
D^{V\mu\nu}(p^2)\sim \frac{ip^\mu p^\nu}{p^2M_V^2}.
\end{equation}
The integral (\ref{loopintegral}) diverges for large loop momenta by reason of the power counting of numerators and denominators in loop graphs.
Introducing a cutoff $\Lambda_C$ violates gauge invariance, Lorentz invariance and unitarity and we find that new more severe divergences in diagrams
containing more loops generate more cutoff parameters, and ultimately an infinite number of unknown parameters appears in the calculation. These
divergences cannot be renormalized and no meaningful predictions can be made in standard local QFT. We note that this is true even though the basic coupling constant $g$ is dimensionless.

The offending factor in the numerator of (\ref{masspropagator}) arises from the spin sum:
\begin{equation}
\label{polarization}
\sum\epsilon^\mu(p,\lambda)\epsilon^{\nu*}(p,\lambda)=-\eta^{\mu\nu}+\frac{p^\mu p^\nu}{M^2_V},
\end{equation}
where the polarization vector $\epsilon^\mu$ has definite spin projection $\lambda=\pm 1,0$ along the $z$-axis, while the $x$- and $y$ directions are
transverse. This corresponds to the three independent polarization vectors for a spin 1 particle. For large values of $p$ the longitudinal state
$\epsilon^\mu(p,\lambda=0)$ is proportional to $p^\mu$, leading to the numerator term $p^\mu p^\nu/M^2_V$ in (\ref{masspropagator}). The {\it raison
d'$\hat{e}$tre} of the spontaneous symmetry breaking Higgs mechanism is the ``gauging away" of the $p^\mu p^\nu/M^2_V$ term in (\ref{masspropagator}) and (\ref{polarization}), making way for a renormalizable EW theory~\cite{thooft} that avoids a violation of unitarity when $\sqrt{s} > 1$ TeV. The new massive boson propagator has the form:
\begin{equation}
\label{tHooftpropagator}
D^{V\mu\nu}(p^2)=\frac{i\biggl(-\eta^{\mu\nu}+\frac{(1-\xi)p^\mu p^\nu}{p^2-\xi M^2_V}\biggr)}{p^2-M^2_V+i\epsilon},
\end{equation}
where $\xi$ is the gauge parameter. The dangerous factor $p^\mu p^\nu/M^2_V$ can now be gauged away by choosing $\xi=1$. For the ``unitary" gauge
$\xi\rightarrow\infty$, the massive spin 1 propagator reverts to (\ref{masspropagator}).

Lee and Yang demonstrated in an early attempt to unify electromagnetism and weak interactions that a non-Abelian gauge invariance of a massive charged vector field can lead to a renormalizable field theory~\cite{LeeYang}. It has been argued that it is difficult to extend the gauge symmetry of massive non-Abelian vector fields without simultaneously losing renormalizability and unitarity~\cite{Salam,Goto,Burnel,Cornwall}.

Let us fix the gauge of the $W_\mu$ and $Z_\mu$ Lagrangians:
\begin{equation}
\label{Wgaugedfixed}
{\cal L}_{\rm WGF}=-\frac{1}{2}W^+_{\mu\nu}W^-_{\mu\nu}+M_W^2W^+_\mu W^{-\mu}
+\frac{1}{\lambda}(D^W_\mu W^{+\mu})(D^W_\nu W^{-\nu}),
\end{equation}
and
\begin{equation}
\label{Zgaugefixed}
{\cal L}_{\rm ZGF}=-\frac{1}{4}Z^{\mu\nu}Z_{\mu\nu}+\frac{1}{2}M_Z^2Z_\mu Z^\mu+\frac{1}{2\kappa}(D^Z_\mu Z^\mu)(D^Z_\nu Z^\nu),
\end{equation}
where $\lambda$ and $\kappa$ are constant gauge parameters.

The equations of motion for the $W^{\pm}$ and $Z$ become
\begin{equation}
D^W_\mu W^{\mu\nu}+M_W^2W^\nu+\frac{1}{\lambda}D^{W\nu}(D^W_\mu W^\mu)=0,
\end{equation}
and
\begin{equation}
D^Z_\mu Z^{\mu\nu}+M_Z^2Z^\nu+\frac{1}{\kappa}D^{Z\nu}(D^Z_\mu Z^\mu)=0.
\end{equation}
The free kinetic energy part of the $W^{\pm}$ equation of motion takes the form:
\begin{equation}
\label{Weqmotion}
(\Box + M_W^2)W^\nu-\partial^\nu(\partial_\mu W^\mu)+\frac{1}{\lambda}\partial^\nu(\partial_\mu W^\mu)=0,
\end{equation}
where $\Box=\partial^\mu\partial_\mu$. A similar equation of motion follows for the $Z$ boson.

The propagator for the $W$ boson consist of two parts: a spin-1 part:
\begin{equation}
-i\frac{\eta_{\mu\nu}-\frac{p_\mu p_\nu}{M^2_W}}{p^2-M^2_W+i\epsilon},
\end{equation}
and a spin-0 part:
\begin{equation}
\label{spin0}
-i\frac{\frac{p_\mu p_\nu}{M^2_W}}{p^2-\lambda M^2_W-i\epsilon},
\end{equation}
We may now suspect that the presence of the spin 0 part can produce a renormalizable theory~\cite{LeeYang}. However, the difference in signs $\pm i\epsilon$ in the two parts of the propagator tells us that this is not the case. The $W$ propagator can be written as
\begin{equation}
\label{tildepropagator}
{\overline D}^W_{\mu\nu}(p^2)=D^W_{\mu\nu}(p^2)-2\pi i\frac{p_\mu p_\nu}{M_W^2}\delta(p^2-\lambda M_W^2),
\end{equation}
where
\begin{equation}
\label{Wpropagator}
D^W_{\mu\nu}(p^2)=-i\biggl(\frac{\eta_{\mu\nu}-\frac{p_\mu p_\nu}{M_W^2}}{p^2-M_W^2+i\epsilon}+\frac{\eta_{\mu\nu}+\frac{p_\mu p_\nu}{M_W^2}}{p^2-\lambda M_W^2+i\epsilon}\biggr).
\end{equation}
The second term in (\ref{tildepropagator}) makes the theory divergent and non-renormalizable. To make the theory renormalizable, we introduce an indefinite metric in the Hilbert space~\cite{LeeYang,Pauli}:
\begin{equation}
{\tilde W}_\mu=\eta^{-1}W_\mu^\dagger\eta,\quad {\tilde Z}=\eta^{-1}Z^\dagger_\mu\eta,
\end{equation}
where $\eta$ represents the indefinite metric. To change the sign of $i\epsilon$ in the spin-0 part of the propagator (\ref{spin0}), the metric $\eta$ is chosen to be
\begin{equation}
\eta=(-1)^{N_s},
\end{equation}
where $N_s$ is the total number of scalar bosons.

Let us consider the free Hamiltonian $H^W_0$ for the $W$ boson field:
\begin{align}
H^W_0={\vec\pi}^W\cdot\tilde{\vec\pi}^W+\lambda\pi^W_0\tilde{\pi}^W_0+M_W^2W_\mu{\tilde W}^\mu+({\vec\nabla}\times{\vec W})\cdot({\vec\nabla}\times\tilde{\vec W})\nonumber\\
+i({\vec\pi}^W{\vec\nabla}W_0+\tilde{\vec\pi}^W{\vec\nabla}{\tilde W}_0-\pi^W_0{\vec\nabla}\cdot{\vec W}-\tilde{\pi}^W_0{\vec\nabla}\cdot\tilde{\vec W}).
\end{align}
Here, $\pi^W_\mu$ is the conjugate momentum operator to $W_\mu$. The equal-time canonical commutation relations are
\begin{equation}
[\pi^W_\mu({\bf x},t),W_\nu({\bf x}',t)]=-i\eta_{\mu\nu}\delta^3({\bf x}-{\bf x}'),\quad [{\tilde\pi}^W_\mu({\bf x},t),{\tilde W}_\nu({\bf x}',t)]=-i\eta_{\mu\nu}\delta^3({\bf x}-{\bf x}').
\end{equation}
All the other equal-time commutators between $W_\mu,\tilde{W_\mu},\pi^W_\mu,{\tilde\pi}^W_\mu$ are zero.

We have
\begin{equation}
\tilde {\vec W}=\eta^{-1}{\vec W}^\dagger\eta,\quad \tilde{\vec\pi}^W=\eta^{-1}{\vec\pi}^{W\dagger}\eta
\end{equation}
and
\begin{equation}
\tilde{W}_0=-\eta^{-1}W_0^\dagger\eta,\quad \tilde{\pi}_0=-\eta^{-1}\pi^{W\dagger}_0\eta.
\end{equation}
In terms of the transverse creation and annihilation operators $a_k^t, b^t_k$, the longitudinal operators $a^l_k, b^l_k$ and the scalar boson operators $a^s_k,b^s_k$, the Hamiltonian $H^W_0$ becomes
\begin{equation}
\label{WHamiltonian}
H^W_0=\sum_{k,t}\omega(a_k^{t\dagger}a^t_k+\frac{1}{2})+\sum_k\omega(a^{l\dagger}_ka_k^l+\frac{1}{2})+\sum_{k,s}\omega_s(a_k^{s\dagger}a_k^s+\frac{1}{2})\\
+{\rm same\,terms\,with}\,\, a\rightarrow b,
\end{equation}
where
\begin{equation}
\omega=({\vec k}^2+M_W^2)^{1/2} > 0,\quad \omega_s=({\vec k}^2+\lambda M_W^2)^{1/2} > 0.
\end{equation}
Moreover, we have
\begin{equation}
\eta=\exp\biggl[\sum_{k,s}i\pi(a_k^{s\dagger}a_k^s+b_k^{s\dagger}b_k^s)\biggr].
\end{equation}
We observe that without the use of the indefinite metric $\eta$ in Hilbert space, the third scalar contribution in (\ref{WHamiltonian}) is negative and the scalar bosons have negative energy.

With the use of an indefinite metric in Hilbert space, the vector boson propagators are given by (\ref{Wpropagator}) and by
\begin{equation}
\label{Zpropagator}
D_{\mu\nu}^Z(p^2)=-i\biggl(\frac{\eta_{\mu\nu}-\frac{p_\mu p_\nu}{M_Z^2}}{p^2-M_Z^2+i\epsilon}+\frac{\eta_{\mu\nu}+\frac{p_\mu p_\nu}{M_Z^2}}{p^2-\kappa M_Z^2+i\epsilon}\biggr).
\end{equation}
A straightforward calculation shows that (\ref{Wpropagator}) and (\ref{Zpropagator}) are equivalent to
\begin{equation}
\label{Wpropagator2}
D^{W\mu\nu}(p^2)=\frac{i\biggl(-\eta^{\mu\nu}+\frac{(1-\lambda)p^\mu p^\nu}{p^2-\lambda M^2_W}\biggr)}{p^2-M^2_W+i\epsilon},
\end{equation}
and
\begin{equation}
\label{Zpropagator2}
D^{Z\mu\nu}(p^2)=\frac{i\biggl(-\eta^{\mu\nu}+\frac{(1-\kappa)p^\mu p^\nu}{p^2-\kappa M^2_Z}\biggr)}{p^2-M^2_Z+i\epsilon},
\end{equation}
which are the same as the propagator (\ref{tHooftpropagator}) in the standard model with a Higgs mechanism~\cite{thooft}. We note that there are scalar
boson poles at $p^2=\lambda M_W^2$ and $p^2=\kappa M_Z^2$. Moreover, the limits $M_W\rightarrow 0$ and $M_Z\rightarrow 0$ can be safely taken in the
propagators (\ref{Wpropagator2}) and (\ref{Zpropagator2}).

The vacuum expectation values of the time ordered products for the $W$ and $Z$ bosons are given by
\begin{equation}
\langle 0\vert T[W_\mu(x)W_\nu(0)]\vert 0\rangle=\frac{1}{(2\pi)^4}\int d^4pD^W_{\mu\nu}(p)\exp(ip\cdot x),
\end{equation}
and
\begin{equation}
\langle 0\vert T[Z_\mu(x)Z_\nu(0)]\vert 0\rangle=\frac{1}{(2\pi)^4}\int d^4pD^Z_{\mu\nu}(p)\exp(ip\cdot x).
\end{equation}
The fermion propagator of our EW theory in momentum space is given by
\begin{equation}
\label{fermionprop}
S=\frac{-i}{\slashed p-m_f+i\epsilon}.
\end{equation}

We observe that for $p^2\rightarrow\infty$ the propagators (\ref{Wpropagator2}) and (\ref{Zpropagator2}) behave like
\begin{equation}
D^{W,Z}(p^2)\sim 1/p^2.
\end{equation}
From the Feynman diagrams for coupling to fermions and from self-couplings of the $W$ and $Z$ bosons, we can prove that the EW theory is renormalizable and unitary, {\it provided the scalar fields are not present in the physical sector of the interactions.}

\section{Indefinite Metric, Unitarity and PT Symmetry}

The Hamiltonian for our EW theory is not Hermitian but satisfies
\begin{equation}
\tilde{H}=\eta^{-1}H^\dagger\eta=H.
\end{equation}
Moreover, the $S$ matrix is not unitary
\begin{equation}
\tilde{S}= \eta^{-1}S^\dagger\eta=S^{-1}.
\end{equation}

In standard quantum field theory the Hamiltonian is Hermitian, $H^\dagger=H$, and we are assured that the energy spectrum is real and positive and that the time evolution of the operator $U=\exp(itH)$ is unitary and probabilities are positive and preserved for particle
transitions. However, in recent years there has been a growth of activity in studying quantum theories with pseudo-Hermitian
Hamiltonians, which satisfy the generalized property of adjointness, $\tilde H=\eta^{-1}H^\dagger\eta$, associated with an
indefinite metric in Hilbert space~\cite{Bender,Bender2}.

Spectral positivity and unitarity can in special circumstances follow from a symmetry property of the Hamiltonian in terms of the
symmetry under the operation of ${\cal P}{\cal T}$, where ${\cal P}$ is a linear operator represented by parity reflection, while
${\cal T}$ is an anti-linear operator represented by time reversal. If a Hamiltonian has an unbroken ${\cal P}{\cal T}$
symmetry, then the energy levels can in special cases be real and the theory can be unitary and free of ``ghosts''. The operation of
${\cal P}$ leads to $\vec{x}\rightarrow -\vec{x}$, while the operation of ${\cal T}$ leads to $i\rightarrow -i$ (or
$x^0\rightarrow -x^0$). Under the operation of ${\cal P}{\cal T}$ the Hamiltonians $H^W$ and $H^Z$ are
invariant under the ${\cal P}{\cal T}$ transformation.

The proof of unitarity follows from the construction of a linear operator ${\cal C}$. This operator is used to define the inner
product of state vectors in Hilbert space:
\begin{equation}
\label{innerprod} \langle\Psi\vert\Phi\rangle=\Psi^{{\cal C}{\cal
P}{\cal T}}\cdot\Phi.
\end{equation}
Under general conditions, it can be shown that a necessary and sufficient condition for the existence of the inner product
(\ref{innerprod}) is the reality of the energy spectrum of $H$~\cite{Bender,Bender2}. With respect to this inner
product, the time evolution of the quantum theory is unitary. In quantum mechanics and in quantum field theory, the operator ${\cal
C}$ has the general form
\begin{equation}
{\cal C}=\exp(Q){\cal P},
\end{equation}
where $Q$ is a function of the dynamical field theory variables. The form of ${\cal C}$ must be determined by solving for the
function $Q$ in terms of chosen field variables and field equations. The form of ${\cal C}$ has been calculated for several
simple field theories, e.g. $\phi^3$ theory and also in massless quantum electrodynamics with a pseudo-Hermitian Hamiltonian. The
solution for ${\cal C}$ satisfies
\begin{equation}
{\cal C}^2=1,\quad [{\cal C},{\cal P}{\cal T}]=0,\quad [{\cal
C},H]=0.
\end{equation}
We shall not attempt to determine a specific generalized charge conjugation operator ${\cal C}$ in the present work.

\section{Unitarity of Scattering Amplitudes}

Let us consider the gauge invariant Lagrangian for the $W_\mu$ sector, which contains a massive gauge invariant part and a ghost scalar field part:
\begin{equation}
\label{GhostLagrangian}
{\cal L}_{W\phi}=-\frac{1}{2}W^+_{\mu\nu}W^{-\mu\nu}+(M_WW^+_\mu-P\partial_\mu\sigma
)(M_WW^{-\mu}-P\partial^\mu\sigma)-{\cal D}^W_\mu\phi{\cal D}^{W\mu}\phi+
\frac{1}{2}\lambda M_W^2\phi^2,
\end{equation}
where $\lambda$ is a gauge parameter and ${\cal D}^W_\mu$ is the covariant derivative operator
\begin{equation}
\label{covariantD}
{\cal D}^W_\mu\phi=(\partial_\mu+igP\partial_\mu\sigma)\phi.
\end{equation}
This reduces to the ordinary partial derivative in the gauge $\sigma=0$, when the scalar field $\phi$ becomes a {\it free decoupled field} as in the Abelian $U(1)$ Stuekelberg formalism. Other gauges can be given by
\begin{equation}
\sigma=f(\phi).
\end{equation}
Such gauges exist because they can be reached by a finite unitary gauge transformation~\cite{Burnel}.

It is assumed that $f(\phi)$ reduces to $\pm\phi$ in lowest orders of an expansion in $\phi$. This guarantees that the propagator $D_{\mu\nu}^W$ is well behaved at high energies. The problems with renormalizability are then confined to the vertices with at least one ghost scalar field. In the case of the Abelian $Z^0$ boson, gauges exist that assure renormalizability.

In standard intermediate vector boson theory unitarity bounds are violated in scattering amplitudes at the level of the tree graph, Born approximation. The unitarity violating processes involve external $W$ bosons. Consider the process $\nu_\mu+\bar{\nu}_\mu\rightarrow W^++W^-$. The amplitude is given by
\begin{equation}
{\cal M}=g^2\epsilon_\mu^{-*}(k_2,\lambda_2)\epsilon_\nu^{+*}(k_1,\lambda_1)\bar{v}(p_2)\gamma^\mu(1-\gamma_5)
\frac{(\slashed{p}_1-\slashed{k}_1+m_\mu)}{(p_1-k_1)^2-m_\mu^2}\gamma^\nu(1-\gamma_5)u(p_1),
\end{equation}
where the $\epsilon^\pm$ are the polarization vectors of the $Ws$. The $\epsilon^{-*}(k_2,\lambda_2)$ is associated with the outgoing $W^-$ with 4-momentum $k_2$ and polarization state $\lambda_2$, while a similar relation holds for $\epsilon^{+*}$ and $W^+$. To get the cross section, we determine $\vert {\cal M}\vert^2$ and sum over the states of the polarization for each of the $Ws$. We have
\begin{equation}
\label{epsilonsum}
\sum_{\lambda=0,\pm 1}\epsilon_\mu(k,\lambda)\epsilon_\nu^*(k,\lambda)=-\eta_{\mu\nu}+\frac{k_\mu k_\nu}{M_W^2}.
\end{equation}
As we recall, in the frame $k^\mu=(k^0,0,0,\vert{\vec k}\vert)$, the momentum dependence is associated with the longitudinal polarization vector $\epsilon^\mu(k,\lambda=0)$ which behaves as $k^\mu/M_W$ at high energies. A calculation shows that the total cross section behaves at large energies as $d\sigma/d\Omega \sim E^2$, which violates partial-wave unitarity for $E\sim 1-2$ TeV~\cite{Aitchison}. The scattering amplitude for the process $W^+_L + W^-_L\rightarrow W^+_L + W^-_L$ is well-known to violate unitarity for $E\sim 1-2$ TeV, due to the high energy behavior of the longitudinal polarization vector for the external $Ws$~\cite{LlewellynSmith,Cornwall}.

We now appeal to the {\it gauge invariance} of our EW theory. Consider a process with the amplitude ${\cal M}_W$ involving an external $Z$. We may write
\begin{equation}
{\cal M}_Z=\epsilon^\mu T_\mu,
\end{equation}
where $T_\mu$ depends on the physical process under consideration. The gauge invariance of our theory implies that if we replace $\epsilon^\mu$ by $k^\mu$, then we must get
\begin{equation}
\label{WardTakahashi}
k^\mu T_\mu=0,
\end{equation}
which is the result of the Ward-Takahashi identity~\cite{Ward}. Thus, as in QED the gauge invariance removes the action of the contribution $k^\mu k^\nu/M_Z^2$ in (\ref{epsilonsum}), and it can be taken to be just $-\eta_{\mu\nu}$ provided that the Ward-Takahashi identity (\ref{WardTakahashi}) holds, a condition that follows from gauge invariance. This result can guarantee that for our EW theory there is no violation of unitarity for scattering amplitudes involving external $Zs$. The key ingredient in the $U(1)$ Abelian case of the $Z$ boson is that the scalar longitudinal field component does not couple to the gauge bosons. 

If we try to repeat this argument for the massive non-Abelian $W$ boson, then we find that the longitudinal scalar field component of the $W$ boson, or the scalar gauge field $\sigma$ appears with interactions which are not renormalizable by power counting. Moreover, the Ward-Takahashi identities do not guarantee unitarity of scattering amplitudes involving external $Ws$, as was the case for the Abelian case. 

If we restrict ourselves to a system of particles with the total energy
\begin{equation}
\label{ScalarEnergy}
E < \mu=\lambda^{1/2}M_W,
\end{equation}
we see that {\it there can be no scalar bosons} in the initial and final states of the scattering matrix. We have that $\mu\rightarrow\infty$ as $\lambda\rightarrow\infty$. This is equivalent to the scalar boson masses becoming infinite. Then, for the initial and final states of the scattering matrix $\eta=+1$ and the S matrix is unitary $S^\dagger=S^{-1}$. For $\lambda$ sufficiently large the scalar boson mass $\mu =\lambda^{1/2}M_W$ can be made big enough to make it difficult to detect the scalar bosons at the LHC. Thus, we are restricted to accept an {\it effective renormalizable theory} which cannot be made UV complete for all energies.

For finite and large values of the scalar boson mass $\mu=\lambda^{1/2}M_W$, such that (\ref{ScalarEnergy}) is satisfied in (\ref{GhostLagrangian}), the theory is gauge invariant and renormalizable and there exist Ward-Takahashi-Slavnov-Taylor identities and a conserved current. In particular, the S-matrix is unitary and because the scalar fields decouple and do not propagate:
\begin{equation}
D_\phi(x-y)=\langle 0\vert T[\phi(x)\phi(y)]\vert 0\rangle=0,
\end{equation}
the scattering amplitudes involving external $Ws$ and longitudinally polarized $Ws$ do not violate unitarity.

In the standard model with a Higgs particle if the Higgs mass $m_H$ gets bigger than a certain value, then perturbation theory will fail and partial-wave unitarity breaks down for the tree level calculation of  $W_L W_L\rightarrow W_L W_L$ scattering. An analysis gives~\cite{Lee}:
\begin{equation}
m_H = \biggl(\frac{8\sqrt{2}\pi}{3G_F}\biggr)^{1/2}\sim 1\,{\rm TeV},
\end{equation}
where $G_F=1.166\times 10^{-5}\,{\rm GeV}^{-2}$ is Fermi's constant determined from muon decay. In our gauge invariant EW theory, the applicability of renormalizable perturbation theory can hold for a much bigger energy for a sufficiently large value of $\mu=\lambda^{1/2}M_W$.

We can replace for $k^2 << M_W^2$ the W-propagator by the constant value $\eta^{\mu\nu}/M_W^2$ and we obtain
\begin{equation}
\frac{G_F}{\sqrt{2}}=\frac{g^2}{8M_W^2}.
\end{equation}
By using $e=g\sin\theta_w$ and adopting the experimental value $\sin^2\theta_w\sim 0.23$, we may then predict the value of $M_W$:
\begin{equation}
M_W=\biggl(\frac{\pi\alpha}{\sqrt{2}G_F}\biggr)^{1/2}\frac{1}{\sin\theta_w}\sim 77.73\,{\rm GeV}.
\end{equation}
We can obtain the $\rho$ parameter measure of the relative strengths of neutral and charged current interactions in four-fermion processes:
\begin{equation}
\rho=\frac{M_W^2}{M_Z^2\cos^2\theta_w}\sim 1
\end{equation}
with
\begin{equation}
M_Z\sim 88.58\, {\rm GeV}.
\end{equation}

\section{Conclusions}

We have constructed an EW theory without explicitly introducing the standard scalar Higgs field kinetic and interaction contributions to the Lagrangian. Instead, we have formulated a Stueckelberg-type EW Lagrangian for the $W$ and $Z$ boson fields. This generates an EW Lagrangian that is gauge invariant and leads to a conserved current. However, the energy of the scalar fields is negative and the scalar fields are ghost particles. To make our gauge invariant massive EW theory physically consistent, we must introduce an indefinite metric in Hilbert space~\cite{Pauli}. This generates a non-Hermitian Hamiltonian and a non-unitary S-matrix. The $W$ and $Z$ boson propagators behave as $1/p^2$ for $p^2\rightarrow\infty$. To reestablish a real positive energy spectrum and a unitary S-matrix, we employ the methods and formalism developed by Bender and collaborators~\cite{Bender,Bender2}. In spite of the existence of the Ward-Takahashi-Slavnov-Taylor identities for scattering amplitudes involving external and longitudinally polarized $Ws$ and a conserved current, the cross sections violate unitarity bounds due to the non-vanishing coupling of the scalar bosons. The scattering amplitudes for longitudinally polarized $Ws$ violate the bound $E^{4-N}$ for large energies $E$ where $N$ denotes the number of external $W$ bosons in the scattering process~\cite{Cornwall}. The electrically neutral massive $Z$ bosons possess a $U(1)$ Abelian gauge invariance and the scalar degrees of freedom associated with the scalar bosons decouple, guaranteeing renormalizability and that unitarity is not violated.

The scalar field boson mass $\mu=\lambda^{1/2}M_W$ can become big for a large value of the parameter $\lambda$. For a big enough $\lambda$ the scalar boson will be very heavy and beyond the current range of detectability of the LHC. For $ E < \mu=\lambda^{1/2}M_W$ the scalar bosons do not enter into the physical scattering matrix. This means that our gauge invariant formulation of EW theory can only be an {\it effective} renormalizable and unitary theory up to a high but finite energy. In the absence of an explicit Higgs spontaneous symmetry breaking, the theory cannot be UV complete to all energies. Unless new physics occurs at some high energy when the scalar fields couple to the matter particles, then a completely renormalizable and unitary EW theory cannot be constructed with the Stuekelberg-type of gauge invariance.

The alternative nonlocal interaction EW theory~\cite{Moffat,Moffat2,Moffat3} predicts scattering amplitudes and the running of the coupling constants $e$, $g$ and $g'$ which will differ significantly from the standard renormalizable theory with a Higgs particle, and the effective renormalizable EW theory we have developed in the present paper. The different scattering amplitudes and cross sections can be distinguished experimentally at the high energies $E > 1-2$ TeV that become accessible at the LHC.

We have left as unknown the origin of fermion and boson particle masses. Because in our model there is no Higgs field, with a spontaneous symmetry breaking of the vacuum pervading spacetime, then the origin of particle masses must be sought in another physical phenomenon.

In the event that the LHC detects a Higgs particle, then the standard EW model can be vindicated. On the other hand, if it is experimentally excluded then we must consider a significantly different EW model in which new fundamental properties of QFT will play a decisive role.

\section*{Acknowledgements}

I thank Viktor Toth, Gerry McKeon and Martin Green for helpful and stimulating discussions. This work was supported by the Natural Sciences and Engineering Research Council of Canada. Research at the Perimeter Institute for Theoretical Physics is supported by the Government of Canada through NSERC and by the Province of Ontario through the Ministry of Research and Innovation (MRI).


\end{document}